\documentclass[preprint,12pt]{aastex}

\slugcomment{Submitted to {\em ApJ}}
\begin{document}
\title{Stochastic Background from Coalescences of NS-NS Binaries}
\author{T. Regimbau}
\affil{UMR 6162 Artemis, CNRS, Observatoire de la C\^ote d'Azur, BP
  4229, 06304 Nice Cedex 4, France}
\email{regimbau@obs-nice.fr}
\and
\author{J. A. de Freitas Pacheco}
\affil{ UMR 6202 Cassiop\'ee, CNRS,Observatoire de la C\^ote d'Azur, B.P.4229, F-06304 Nice Cedex 4, France}
\email{pacheco@obs-nice.fr}

\begin{abstract}
In this work, numerical simulations were used to investigate the gravitational
stochastic background produced by coalescences occurring up to $z \sim 5$ 
of double neutron star systems. The cosmic coalescence rate was derived
from Monte Carlo methods using the probability distributions for forming
a massive binary and to occur a coalescence in a given redshift.
A truly continuous background is produced by events located only beyond
the critical redshift $z_* = 0.23$. Events occurring in the redshift
interval $0.027<z<0.23$ give origin to a ``popcorn" noise, while those arising
closer than $z = 0.027$ produce a shot noise. The gravitational density
parameter $\Omega_{gw}$ for the continuous background reaches a maximum
around 670 Hz with an amplitude of $1.1\times 10^{-9}$, while the ``popcorn"
noise has an amplitude about one order of magnitude higher and the
maximum occurs around a frequency of 1.2 kHz. 
The signal is below the sensitivity of the first generation of
detectors but could be detectable by the future generation of ground
based interferometers. Correlating two coincident advanced-LIGO detectors
or two EGO interferometers, the expected S/N ratio are respectively
0.5 and 10.

\end{abstract} 


\section{Introduction}

The merger of two neutron stars, two black holes or a 
black hole and a neutron star are among the most important sources
of gravitational waves (GW), due to the huge energy released in the
process. In particular, the coalescence of double neutron stars (DNS) may radiate
about 10$^{53}$ erg in the last seconds of their inspiral trajectory,
at frequencies up to 1.4-1.6 kHz, range covered by most of the
ground-based laser interferometers like VIRGO \citep{bra},
LIGO \citep{abr}, GEO \citep{hou} or TAMA \citep{kur}. Besides the amount
of energy involved in these events, the rate at which they occur in the local
universe is another parameter characterizing if these mergings are or not potential
interesting sources of GW. In spite of
the large amount of work performed in the past years, uncertainties
persist in estimates of the DNS coalescence rate. In a previous 
investigation, we have revisited this question \citep{dfp,reg05}, taking
into account the galactic star formation history derived directly
from observations and including the contribution of elliptical galaxies
when estimating the mean merging rate in the local universe. Based on
these results, we have predicted a detection rate of one event every 125 and 148 years  
by initial LIGO and VIRGO respectively and up to 6 detections per year in their 
advanced configurations.

Besides the emission produced by the coalescence of the nearest DNS, the 
superposition of a large number of unresolved sources at high redshifts will
produce a stochastic background of GW.
In the past years, different astrophysical processes susceptible to generate a stochastic 
background have been investigated. On the one hand, distorted black holes
\citep{fer99,ara}, bar mode emission from young
neutron stars \citep{reg} are examples of sources able to generate
a shot noise (time interval between events large in comparison with
duration of a single event), while supernovas or hypernovas
\citep{bla,cow01,cow02,buo} are expected to produce an
intermediate ``popcorn" noise. On the other hand, the contribution of
tri-axial rotating neutron stars \citep{reg01}, including magnetars \citep{rp05},
constitutes a truly continuous background. 

Populations of compact binaries as, for instance, the cataclismic variables are
responsible for the existence of a galactic background of GW in the mHz domain,
which could represent an important source of confusion noise for space
detectors as LISA \citep{eva,hils,ben,pos,nel,tim}.
These investigations have been extended recently to the extra-galactic contribution. 
\citet{sch}, \citet{kim04} and \citet{coo} considered cosmological
populations of double and mixed systems involving black holes, neutron stars and
white dwarfs, while close binaries originated from low and intermediate mass stars
were discussed by \citet{far}. 

In this work, using the DNS merging rate estimated in our precedent study, we
have estimated the gravitational wave background spectrum produced by these
coalescences. Numerical simulations based on Monte Carlo methods were performed in order to 
determine the critical
redshift $z_*$ beyond which the duty cycle condition required to have a continuous
background ($D > 1$) is satisfied. Unlike previous studies which focus their attention on 
the early low frequency inspiral phase covered by LISA \citep{sch,far,coo}, here we are
mainly interested in the few thousand seconds before the last stable orbit is reached, 
when more than 96\% of the gravitational wave energy is released. The signal frequency is in
the range 10-1500 Hz, covered by ground based interferometers.
The paper is organized as follows. In \S2, the simulations are described;
in \S3 the contribution of DNS coalescences to
the stochastic background is calculated; in \S4 the detection possibility
with laser beam interferometers is discussed and, finally, in \S5 the main 
conclusions are summarized.

\section{The Simulations}

In order to simulate by Monte Carlo methods the occurrence of merging events, we have adopted 
the following procedure. The first step was to estimate the probability for a given pair
of massive stars, supposed to be the progenitors of DNS, be formed at a
given redshift. This probability distribution is essentially given by
the cosmic star formation rate  \citep{cow02}, normalized in the redshift interval 
$0 \leq z \leq 5$, e.g., 
\begin{equation}
P_f(z)=\frac{dR_f(z)/dz}{N_{p}}
\label{eq-fproba}
\end{equation}
The normalization factor in the denominator is essentially the rate at which massive binaries
are formed in the considered redshift interval, e.g.,
\begin{equation}
N_{p}=\int_0^5(dR_f(z)/dz)dz,
\label{eq-Nevents}
\end{equation}
which depends on the adopted cosmic star formation rate, as we shall see later.

The formation rate of massive binaries per redshift interval is
\begin{equation}
R_{zf}(z_f)=\frac{dR_f(z_f)}{dz_f }=\lambda_{p} \frac{R^*_{f}(z_f)}{1+z_f}{\frac{{dV(z_f)}}{{dz_f}}}
\label{eq-frate}
\end{equation}
In the equation above, $R^*_{f}(z)$ is the cosmic star formation rate (SFR)
expressed in M$_{\odot}$ Mpc$^{-3}$yr$^{-1}$ and $\lambda_{p}$ is the
mass fraction converted into DNS progenitors. 
Hereafter, rates per comoving volume will always be indicated
by the superscript '*', while rates with  indexes ``z$_f$" or ``z$_c$" refer to 
differential rates per redshift interval, including all cosmological factors.
The (1+z) term in the denominator of eq.~\ref{eq-frate} corrects the star formation rate by time 
dilatation due to the cosmic expansion. In the present work we assume that the parameter
$\lambda_p$ does not change significantly with the redshift and thus it will be considered as a constant.
In fact, this term is the product of three other parameters, namely,
\begin{equation}
\lambda_{p}= \beta_{NS} f_b \lambda_{NS}
\end{equation}
where $\beta_{NS}$ is the fraction of binaries which remains
bounded after the second supernova event, $f_b$ is the fraction of massive binaries formed among all
stars and $\lambda_{NS}$ is the mass fraction of neutron star progenitors. 

According to the results by \citet{dfp},\citet{reg05}, $\beta_{NS}$ = 0.024 and $f_b$ = 0.136, values which
will be adopted in our calculations. Assuming that progenitors with initial masses above 40 M$_{\odot}$ 
will produce black holes and considering an initial mass function (IMF) of the 
form $\xi(m) = Am^{-\gamma}$, with $\gamma$ = 2.35 \, (Salpeter's law),
normalized within the mass interval 0.1 - 80 M$_{\odot}$ such as $\int
m\xi(m)dm$ = 1, it results finally $\lambda_{NS} = \int_9^{40}\xi(m)dm = 5.72~\times
10^{-3}~ M_{\odot}^{-1}$ and $\lambda_{p}=1.85 \times
10^{-5}$~M$_{\odot}^{-1}$. 
The evaluation of the parameters $\beta_{NS}$ and $f_b$ depends on different assumptions, which
explain why estimates of the coalescence rate of DNS found in the literature may vary by one or even
two orders of magnitude. The evolutionary scenario of massive binaries considered in our calculations
(see \citet{dfp} for details) is similar to that developed by \citet{bel}, in which none of the stars
ever had the chance of being recycled by accretion. Besides the evolutionary path, the resulting fraction 
$\beta_{NS}$ of bound NS-NS binaries depends on the adopted velocity distribution of the natal kick.
The imparted kick may unbind binaries which otherwise might have remained bound or, less probably, conserve
bound systems which without the kick would have been disrupted. The adopted value for $\beta_{NS}$
corresponds to a 1-D velocity dispersion of about 80 km/s. This value is smaller than those usually
assumed for single pulsars, but consistent with recent analyses of the spin period-eccentricity
relation for NS-NS binaries \citep{dewi}. Had we adopted a higher velocity dispersion in our
simulations (230 km/s instead of 80 km/s), the resulting fraction of bound systems is reduced
by one order of magnitude, e.g., $\beta_{NS}$ = 0.0029. If, on the one hand the
fraction of bound NS-NS systems after the second supernova event depends on the previous evolutionary
history of the progenitors and on the kick velocity distribution, on the other hand estimates of the
fraction $f_b$ of massive binaries formed among all stars depends on the ratio between single and
double NS systems in the Galaxy and on the value of $\beta_{NS}$ itself \citep{dfp}. We have
estimated relative uncertainties of about $\sigma_{f_b}/f_b \approx 0.5$ and 
$\sigma_{\beta_{NS}}/\beta_{NS} \approx 0.75$, leading to a relative uncertainty in the parameter
$\lambda_p$ of about $\sigma_{\lambda_p}/\lambda_p \approx 0.9$. However, we emphasize that these
are only formal uncertainties resulting from our simulations, which depend on the adopted
evolutionary scenario for the progenitors. A comparison with other estimates can be found
in \citet{dfp}.

The element of comoving volume is given by
\begin{equation}
dV(z) = 4\pi r(z)^2 \frac{c}{H_o} \frac{dz}{E(\Omega_i,z)}
\end{equation}
with
\begin{equation}
E(\Omega_i,z) =[\Omega_m(1+z)^3+\Omega_v]^{1/2}
\end{equation}
where $\Omega_m$ and $\Omega_v$ are respectively the present values of
the density parameters due to matter (baryonic and non-baryonic) and
vacuum, corresponding to a non-zero cosmological constant. 
A ``flat" cosmological model ($\Omega_m + \Omega_v = 1$) was assumed. 
In our calculations, we have taken $\Omega_m$ = 0.30 and $\Omega_v$ = 0.70,
corresponding to the so-called ``concordance" model derived from
observations of distant type Ia supernovae \citep{schm} and the power spectra of the cosmic microwave 
background fluctuations \citep{spe}. The Hubble parameter H$_0$ was taken to be
$65\,\mathrm{km\,s}^{-1}\mathrm{Mpc}^{-1}$.

\begin{figure}
\plotone{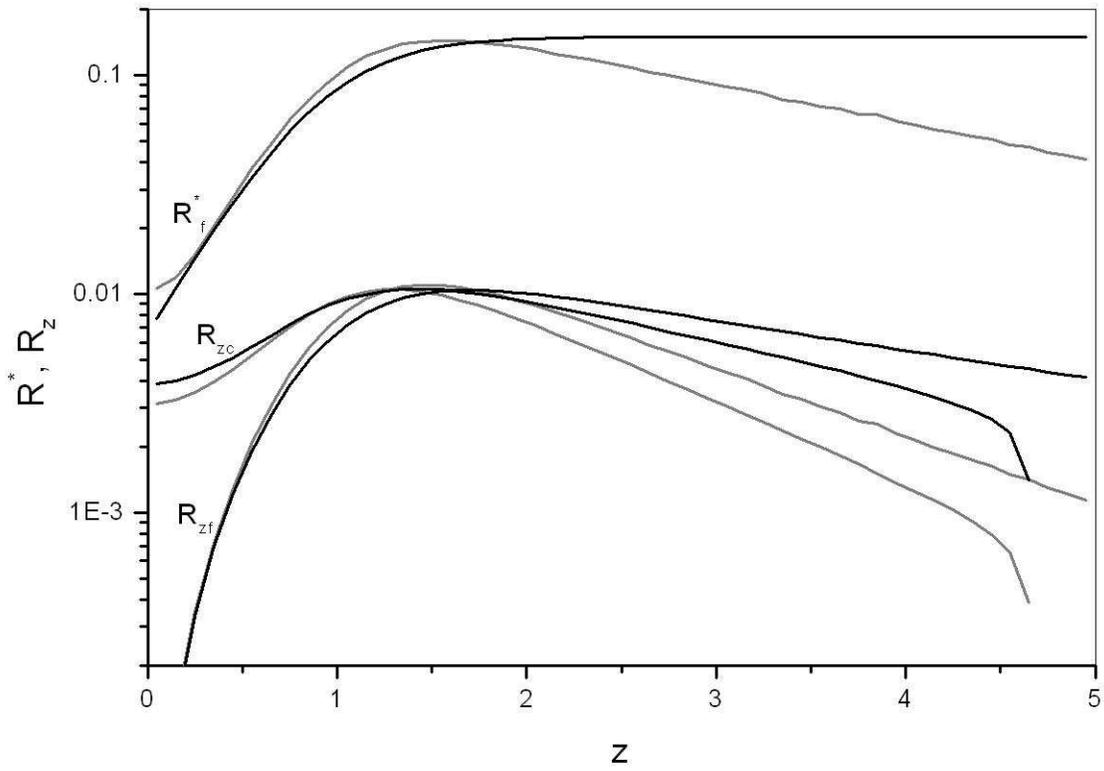}
\caption{
Cosmic star formation rate $R^{*}_f(z)$ in
M$_{\odot}$Mpc$^{-3}$yr$^{-1}$ in a flat cosmology with $\Omega_m$ = 0.30
and $\Omega_v$ =0.70: SFR2 is represented by
a black line and SFR1 by a grey line.
The formation and the coalescence rate per redshift
interval are also plotted for comparison.
\label{fig-sfr}} 
\end{figure}

\citet{por} provide three models for the cosmic SFR
history up to redshifts $z\sim 5$. 
Differences among these models are mainly due to various corrections
applied, in particular those due to extinction by the cosmic dust.
In our computations, we have considered the second model, labelled
SFR2 \citep{mad} but numerical results using SFR1 \citep{ste} will be
also given for comparison. Both rates increase rapidly between $z\sim 0-1$ and peak 
at $z\sim 1-2$, but SFR1 decreases gently after $z\sim 2$ while SFR2 remains more or less 
constant (Figure~\ref{fig-sfr}).

 
The next step consists to estimate the redshift $z_b$ at which the progenitors
have already evolved and the system is now constituted by two neutron stars.
This moment fixes also the beginning of the inspiral phase. If $\tau_b$
($\approx 10^8$ yr) is the mean lifetime of the progenitors (average weighted by the 
IMF in the interval 9-40 M$_{\odot}$) then

\begin{equation}
z_b = z_f - H_0 \tau_b (1+z_f)E(z_f)
\label{eq-zb}
\end{equation}


Once the beginning of the inspiral phase is established, the redshift at which the coalescence
occurs is estimated by the following procedure. The duration of the inspiral phase depends on the 
orbital parameters just after the second supernova and on the neutron star masses. The probability 
for a given DNS system to coalesce in a timescale
$\tau$ was initially derived by \citet{pa97}, confirmed by subsequent simulations 
\citep{vin,dfp} and is given by

\begin{equation}
P_{\tau}(\tau)=B / \tau
\label{eq-proba_tau}
\end{equation}
 
\begin{figure}
\plotone{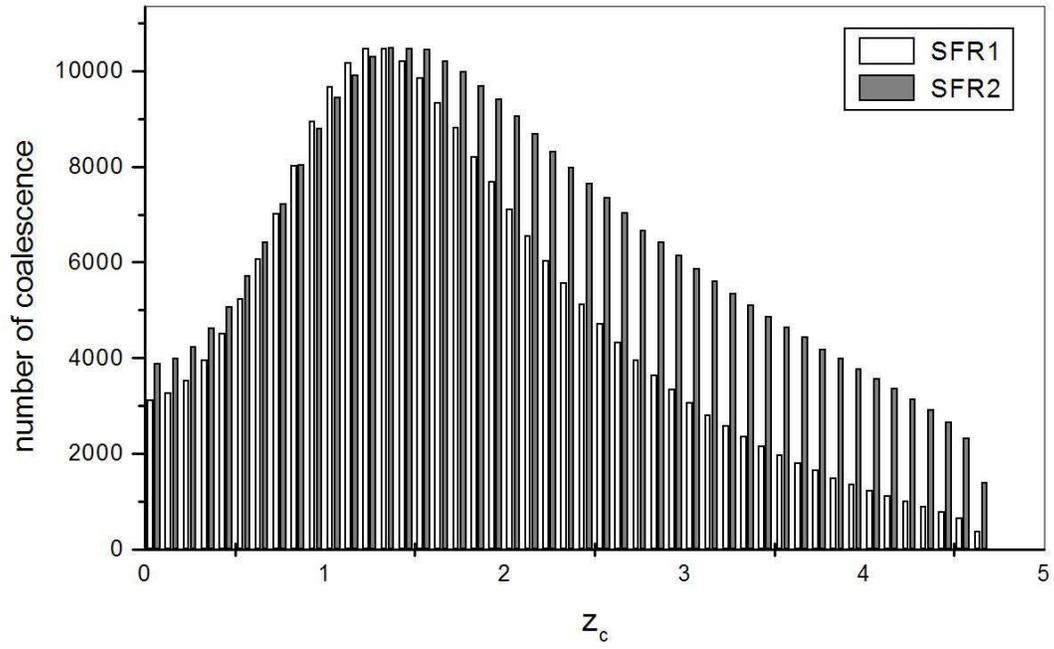}
\caption{
Distribution of coalescences as a function of the redshift derived from our
simulations, corresponding to $10^7$ numerical experiments.
\label{fig-dist_coalescence}} 
\end{figure}

Simulations indicate a minimum coalescence timescale $\tau_0 = 2 \times 10^5$ yr but a considerable 
number of systems have a coalescence timescale higher than the Hubble time. The normalized
probability in the range $2\times 10^5$ yr up to 20 Gyr implies $B=0.087$. Therefore,
the redshift $z_c$ at which the coalescence occurs after a timescale $\tau$ is derived from the equation
\begin{equation}
H_0\tau = \int_{z_c}^{z_b} \frac{dz}{(1+z)E(z)}
\label{eq-H0_tau}
\end{equation}
which was solved in our code by an iterative method. 
The resulting distribution of the number of coalescences as a function of $z_c$ is shown in
Figure~\ref{fig-dist_coalescence} for both star formation rates SFR1 and SFR2, while the
corresponding coalescence rate per redshift interval, $R_{zc}(z)$, is
shown in Figure~\ref{fig-sfr}. In the same figure, for comparison, we have plotted
the formation rate $R_{zf}(z)$ (eq.~\ref{eq-frate}). Notice that the maximum of $R_{zc}(z)$ is shifted
towards lower redshifts with respect to the maximum of $R_{zf}(z)$,
reflecting the time delay between the formation of the progenitors and
the coalescence event. The coalescence rate $R_{zc}(z)$, does not fall to
zero at $z=0$ because a non negligible fraction of coalescences ($\sim
3\%$ for SFR2 and $\sim 5\%$ for SFR1) occurs later than $z=0$.

\section{The Gravitational Wave Background}
 
The nature of the background is determined by the
duty cycle defined as the ratio of the typical duration of a single
burst $\bar{\tau}$ to the average time interval between successive events, e.g.,
\begin{equation}
D(z_*)=\int_{0}^{z_*} \bar{\tau} (1+z')R_{zc}(z')dz'
\label{eq-dc}
\end{equation}
The critical redshift $z_*$ at which the background becomes
continuous is fixed by the condition $D(z)>1$.
Since we are interested in the last instants of the inspiral, when the signal is within
the frequency band of ground based interferometers, we took
$\bar{\tau} = 1000$ s, duration which
includes about $96\%$ of the total energy released (see Table~\ref{tbl-duration}).
From our numerical experiments and imposing D = 1, one obtains $z_* =
0.23$ when the SFR2 is used and $z_* = 0.27$ for the SFR1. About 96\% (94\% in the case of SFR1) of 
coalescences occur above such a redshift, contributing to the production of a 
continuous background. Sources in the redshift interval $0.027 < z <
0.23$ (SFR2) or $0.032 < z < 0.27$ (SFR1) correspond to
a duty cycle D = 0.1 and they are responsible for a cosmic ``popcorn" noise.

The gravitational fluence (given here in  erg cm$^{-2}$Hz$^{-1}$) in the observer 
frame produced by a given DNS coalescence is:

\begin{equation}
f_{\nu _{o}}=\frac{1}{4\pi d_{L}^{2}}\frac{dE_{gw}}{d\nu }(1+z_c)
\label{eq-fluence}
\end{equation}
where $d_{L}=(1+z_c)r$ is the distance luminosity, $r$
is the proper distance, which depends on the adopted cosmology,
${dE_{gw}}/{d\nu}$ is the gravitational spectral energy and
$\nu=(1+z_c)\nu _{o}$ the frequency in the source frame. 
In the quadrupolar approximation and for a binary system with
masses $m_1$ and $m_2$ in a circular orbit: 
\begin{equation}
dE_{gw}/{d\nu} = K \nu^{-1/3}
\end{equation}
where the fact that the gravitational wave frequency is twice the orbital frequency was 
taken into account. Then
\begin{equation}
K = \frac{(G \pi)^{2/3}}{3} \frac{m_1m_2}{(m_1+m_2)^{1/3}}
\end{equation}
Assuming $m_1 = m_2 = 1.4$ one obtains $K=5.2 \times 10^{50}$ erg Hz$^{-2/3}$.


The spectral properties of the stochastic background are characterized
by the dimensionless parameter \citep{fer99}:

\begin{equation}
\Omega _{gw}(\nu _{o})=\frac{1}{c^{3} \rho _{c}}{\nu _{o}}F_{\nu _{o}}
\end{equation}
where $\nu _{o}$ is the wave frequency in the observer frame, $\rho _{c}$ is the critical mass 
density needed to close the Universe, related to the Hubble parameter $H_{0}$ by,

\begin{equation}
\rho _{c}=\frac{3H_{o}^{2}}{8\pi G}
\end{equation}
$F_{\nu_o}$ is the gravitational wave flux (given here in  erg
cm$^{-2}$Hz$^{-1}$s$^{-1}$) at the 
observer frequency
$\nu_o$, integrated over all sources at redshifts $z_c > z_*$, namely
\begin{equation}
F_{\nu _{o}}=\int_{z_*}^{z_{\max }}f_{\nu _{o}}dR_c(z)
\end{equation} 

Instead of solving analytically the equation above by introducing, for instance, an
adequate fit of the cosmic coalescence rate, we have calculated the integrated
gravitational flux by summing 
individual fluences (eq.~\ref{eq-fluence}), scaled by the ratio between the
expected number of events per unit of time and the number of simulated
coalescences or, in other words, the ratio between the  total formation rate of progenitors
(eq.~\ref{eq-Nevents}) and the number of simulated massive binaries, e.g.,

\begin{equation}
F_{\nu_o}=\frac{N_{p}}{N_{sim}}\sum_{i=1}^{N_{sim}} f_{\nu_o}^i 
\end{equation}

The number of runs (or $N_{sim}$) in our simulations was equal to $10^7$, representing
an uncertainty of $\leq$ 0.1\% in the density parameter $\Omega_{gw}$. Using the SFR2, the derived formation rate of progenitors is
$N_p = 0.031~s^{-1}$, whereas for the SFR1 one obtains $N_p = 0.024~s^{-1}$.
For each run,
the probability distribution $P_f(z)$ defines, via Monte Carlo, the redshift at
which the massive binary is formed. The beginning of the inspiral phase at $z_b$ is fixed
by eq.~\ref{eq-zb}. Then, in next step, the probability distribution of the coalescence timescale
and eq.~\ref{eq-H0_tau} define the redshift $z_c$ at which the merging occurs. The fluence produced
by this event is calculated by eq.~\ref{eq-fluence}, stored in different frequency bins in the
observer frame and added according to the equation above. 

\begin{figure}
\plotone{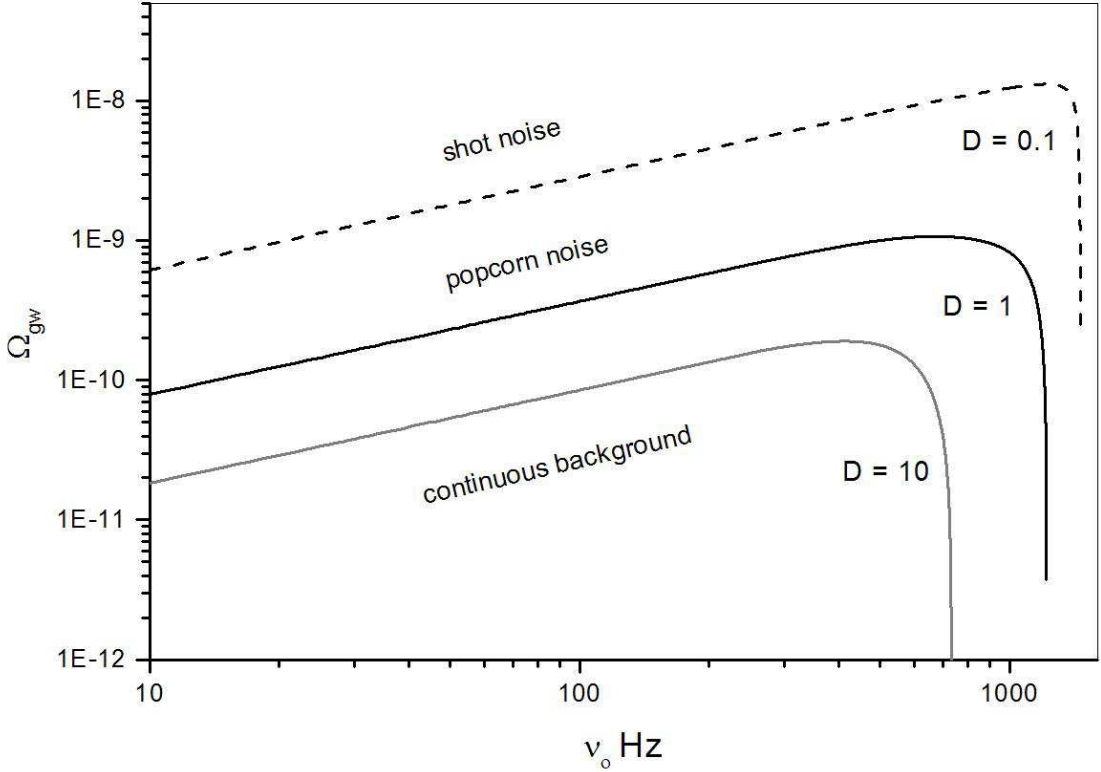}
\caption{Spectrum of the expected gravitational energy density
  parameter $\Omega_{gw}$ corresponding to NS-NS coalescences
  occurring beyond $z_* = 0.23$ (bold continuous curve). A more
  conservative background ($D>10$) corresponding to sources beyond
  $z=1.05$ is plotted for comparison (grey continuous curve). The
  expected ``popcorn" noise contribution arising from sources in the
  redshift interval $0.027<z<0.23$ (dashed curve) is also shown. This
  corresponds to the intermediate zone between a shot noise ($D<0.1$) and a continuous background ($D>1$).\label{fig-omega}}
\end{figure}

Figure ~\ref{fig-omega} shows the density parameter $\Omega_{gw}$ as a function of
the observed frequency derived from our simulations.
The density parameter  $\Omega_{gw}$ increases as $\nu_{o}^{2/3}$ at low frequencies, reaches a maximum 
amplitude of about $1.1 \times 10^{-9}$ around 670 Hz in the case of SFR2 and a maximum of $8.4 \times 10^{-10}$) around 630 Hz, in the case of SFR1.  
A high frequency cut-off at $\sim 1220$ Hz ($\sim 1170$ Hz for SFR1 ) is observed, corresponding 
approximately to the frequency of the last stable orbit at the critical redshift $z_*=0.23$ 
($z_*=0.27$ for SFR1).
Calculations performed by \citet{sch}, in spite of the similar local
merging rates, indicate that the maximum occurs at lower
frequencies ($\sim$ 100 Hz) with an amplitude (scaled to the Hubble parameter
adopted in this work) lower by a factor of seven. However, as those authors have stressed, their 
calculations are expected to be accurate in the frequency range 10 $\mu$Hz up to 1 Hz, since
they have set the value of the maximum frequency $\nu_{max}$ to about that expected at
a separation of three times the ``last stable orbit" (LSO), e.g.,
$\nu_{max} \approx 0.19\nu_{LSO}$.
Thus, a direct comparison with our results is probably meaningless. 

A more conservative estimate can be obtained if one adopts a higher duty cycle value, namely, $D > 10$,
corresponding to sources located beyond  $z \sim 1.05$ ($z \sim
1.075$ for SFR1). Our results are plotted in Fig.~\ref{fig-omega},
which includes, for comparison, the ``popcorn" noise contribution
arising from sources between $0.027 < z < 0.23$ ($0.032 < z < 0.27$ for SFR1), corresponding to the
intermediate zone between a shot noise ($D<0.1$) and a continuous
background ($D>1$). When increasing the critical redshift (or
removing the nearest sources), the
amplitude of $\Omega_{gw}$ decreases and the spectrum is shifted
toward lower frequencies.
The amplitude of the ``popcorn" background is about one order of magnitude
higher than the continuous background, with a maximum of about
$\Omega_{gw}=1.3 \times 10^{-8}$ ($8.8 \times 10^{-9}$ for SFR1) around a frequency of $1.2$ kHz. 

\begin{figure}
\plotone{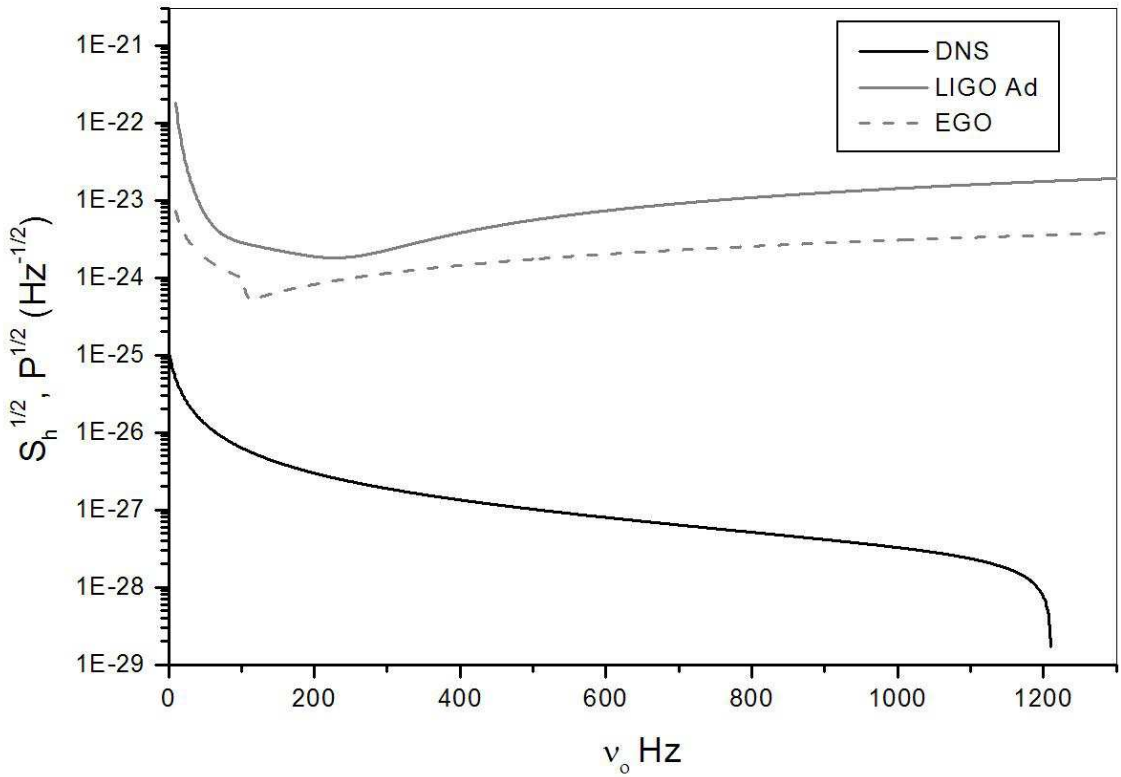}
\caption{Gravitational strain in Hz$^{-1/2}$ corresponding to NS-NS
  coalescences occurring beyond $z_* = 0.23$, along with the planned sensitivity
  curves of LIGO Ad (continuous curve) and EGO (dashed curve).\label{fig-Sh}}
\end{figure}

Since some authors use, instead of $\Omega_{gw}$, the gravitational strain $\sqrt{S_h}$
defined by \citet{all99} as 
\begin{equation}
S_h(\nu_{o})=\frac{3H_{0}^2}{10 \pi^2} \frac{1}{\nu_{o}^3}\Omega_{\rm gw}(\nu_{o})
\end{equation} 
we show this quantity in Figure~\ref{fig-Sh}.

\section{Detection}

Because the background obeys a Gaussian statistic and can be confounded with
the instrumental noise background of a single detector, the optimal detection 
strategy is to cross-correlate the output of two (or more) detectors, assumed to have 
independent spectral noises. The cross correlation product is given by \citep{all99}:

\begin{equation} 
Y=\int_{-\infty}^\infty \tilde{s_1}^*(f)\tilde{Q}(f)\tilde{s_2}(f) df
\end{equation}
where
\begin{equation}
\tilde{Q}(f)\propto \frac{\Gamma (f) \Omega_{gw}(f)}{f^3P_1(f)P_2(f)}
\end{equation}
is a filter that maximizes the signal to noise ratio ($S/R$). In the
above equations, $P_1(f)$ and  $P_2(f)$ are the power spectral noise
densities of the two detectors and $\Gamma$ is the non-normalized overlap
reduction function, characterizing the loss of sensitivity due to
the separation and the relative orientation of the detectors. 
The optimized $S/N$ ratio for an integration time $T$ is given by \citep{all97} :
\begin{equation}
(\frac{S}{N})^2 =\frac{9 H_0^4}{8 \pi^4}T\int_0^\infty
df\frac{\Gamma^2(f)\Omega_{gw}^2(f)}{f^6 P_1(f)P_2(f)}.
\label{eq-snr}
\end{equation} 

In the literature, the sensitivity of detector pairs is usually given in terms of the minimum
detectable amplitude for a flat spectrum ($\Omega_{gw}$ equal to constant) \citep{all99}, e.g.,

\begin{equation}
\Omega_{min}=\frac{4 \pi^2}{3H_0^2\sqrt{T}}({\rm erfc}^{-1}(2 \alpha)-{\rm erfc}^{-1}(2 \gamma)) 
 \lbrack \int_0^\infty df \frac{\Gamma^2 (f)}{f^6 P_1(f) P_2(f)}\rbrack^{-1/2}
 \end{equation}

The expected minimum detectable amplitude for the main pair of detectors  in the world, after one
year integration, are given in Table~\ref{tbl-sensitivity}, for a detection rate $\alpha=90\%$  and a 
false alarm rate $\gamma=10\%$.
The power spectral densities expressions used for the present
calculation can be found in \citet{dam}.
$\Omega_{min}$ is of the order of $10^{-6}-10^{-5}$ for the first generation of interferometers
combined as LIGO/LIGO and LIGO/VIRGO. Their advanced counterparts will permit an increase of
two or even three orders of magnitude in sensitivity ($\Omega_{min} \sim 10^{-9}-10^{-8}$).
The pair formed by the co-located and co-aligned LIGO Hanford detectors, for
which the overlap reduction function is equal to one, 
is potentially one order of magnitude more sensitive than the
Hanford/Livingston pair, provided that instrumental and environmental
noises could be removed.

\begin{figure}
\plotone{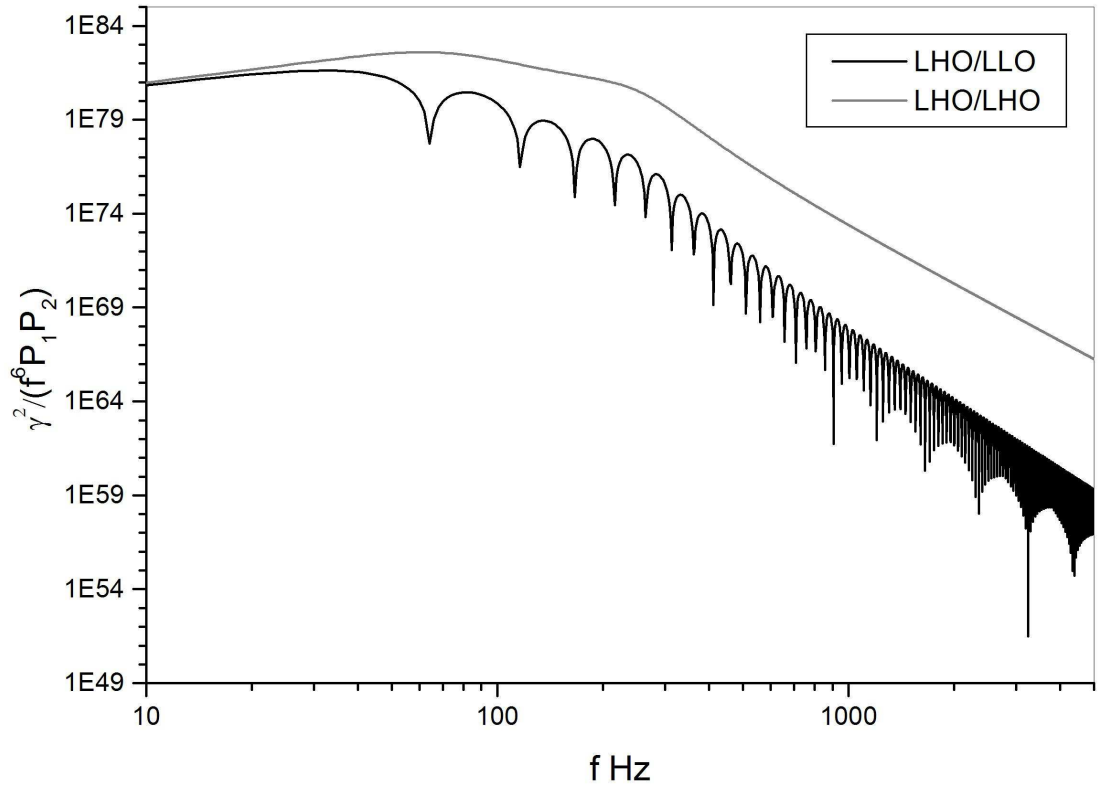}
\caption{Integrand of the signal-to-noise ratio (eq.~\ref{eq-snr}) for advanced LIGO
pairs: LHO-LHO (grey curve) and LHO-LLO (black curve)\label{fig-snr}}
\end{figure}

However, because the spectrum of DNS coalescences {\it is not flat} and the maximum occurs
out of the optimal frequency band of ground based interferometers,
which is typically around $50-300$ Hz, as shown in
Figure~\ref{fig-snr}, the S/N ratio is slightly reduced.
Considering the co-located and co-aligned LIGO interferometer
pair, we find a signal-to-noise ratio of $S/N \sim 0.002$ ($S/N \sim
0.5$) for the
initial (advanced) configuration.
Unless the coalescence rate be substantially higher than
the present expectations, our results indicate that their contribution to the
gravitational background is out of reach of the first and the second
generation of interferometers.
On the other hand, the sensitivity of the future third generation of
detectors, presently in discussion, could be high enough to gain one
order of magnitude in the expected S/N ratio.
Examples are the Large Scale Cryogenic Gravitational
Wave Telescope (LCGT), sponsored by the University of Tokyo and the
European antenna EGO (Sathyaprakash, private communication).
EGO will incorporate signal recycling, diffractive optics on silicon
mirrors, cryo-techniques and kW-class lasers, among other
technological improvements.
A possible sensitivity for this detector is shown in Figure~\ref{fig-Sh}, compared 
to the expected sensitivity of advanced LIGO.
Around 650 Hz, the planned strain noise $\sqrt{S_n(\nu)}$ is about $8
\times 10^{-24}$ Hz$^{-1/2}$ for the advanced LIGO configuration while at this
frequency, the planned strain noise for EGO is $2\times 10^{-24}$
Hz$^{-1/2}$, which represents a gain by a factor of $\sim 4$. 
Considering two interferometers located at the same place, we find a signal-to-noise 
ratio $S/N \sim 10$.

On the other hand, the popcorn noise contribution  
could be detected by new data analysis techniques currently under
investigation, such as the search for anisotropies \citep{all} that can be used to create a 
map of the GW background \citep{cor}, the maximum likelihood statistic \citep{dra}, 
or methods based on the Probability ``Event Horizon" concept
\citep{cow05}, which describes the evolution, as a function of the observation
time, of the cumulated signal throughout the Universe.
The PEH of the GW signal evolves fastly from contributions of high
redshift populations, forming a real continuous stochastic background, to low redshift
and less probable sources that can be resolved individually, while the
PEH of the instrumental noise is expected to evolve much slower.
Consequently, the GW signature could be distinguished from the instrumental noise background.

\section{Conclusions}

In this work, we have performed numerical simulations using Monte Carlo
techniques to estimate the occurrence of double neutron star coalescences
and the gravitational stochastic background produced these events.
Since the coalescence timescale obeys a well defined probability 
distribution ($P(\tau) \propto 1/\tau$), derived from simulations of
the evolution of massive binaries \citep{dfp}, the cosmic coalescence rate does
not follow the cosmic star formation rate and presents necessarily a time-lag.
In the case where the sources are supernovae or black holes, the gravitational burst
is produced in a quite short timescale after the the formation of the progenitors. Therefore,
the time-lag is negligible and the comoving volume where the progenitors are formed
is practically the same as that where the gravitational wave emission occurs, introducing
a considerable simplification in the calculations. This is not the case when NS-NS coalescences
are considered, since timescales comparable or even higher than the Hubble timescale have
non negligible probabilities.
The maximum probability to form a massive binary
occurs at $z \sim 1.7$, depending slightly on the adopted cosmic star formation rate,
whereas the maximum probability to occur a coalescence is around $z
\sim 1.4$.

We have found that a truly continuous background is formed only when sources
located beyond $z > 0.23$ ($z > 0.27$ for the SFR1 case), including 96\% (94\% for SFR1) of all events and the
critical redshift corresponds to the condition $D > 1$. Sources in the
redshift interval $0.027 < z < 0.23$  ($0.032 < z < 0.27$ for SFR1) produce a ``popcorn" noise.
Our computations indicate that the density parameter $\Omega_{gw}$ has
a maximum around 670 Hz (630 Hz for SFR1), attaining an amplitude of about of
$1.1 \times 10^{-9}$ ($8.3 \times 10^{-10}$ for SFR1). The low frequency cutoff around 1.2 kHz corresponds
essentially to the gravitational redshifted wave frequency associated to last stable orbit of sources
located near the maximum of the coalescence rate.
 
The computed signal is below the sensitivity of the first and the
second generation of detectors.
However, using the planned sensitivity of third generation interferometers, we found that after one year of 
integration, the cross-correlation of two EGO like
coincident antennas, gives an the optimized signal-to-noise of $S/R \sim 10$.

The ``popcorn" contribution is one order of magnitude higher with a
maximum of $\Omega_{gw} \sim 1.3 \times 10^{-8}$ ($8.8 \times 10^{-9}$ for SFR1) at $\sim 1.2$ kHz.
This signal, which is characterized by the spatial and temporal
evolution of the events as well as by its signature, can be
distinguished from the instrumental noise background and adequate data
analysis strategies for its detection are currently under investigation \citep{all,cor,dra,cow05}.

{\bf Acknowledgement}

The authors thanks the referee by his useful comments, which have improved the early version
of this paper.

\clearpage
\begin{deluxetable}{cccc}
\tabletypesize{\scriptsize}
\tablecaption{For three values of the emission frequency (column 1), the corresponding time left to the last 
stable orbit (second column) 
and percentage of energy released (third column)\label{tbl-duration}}
\tablewidth{0pt}
\tablehead{
\colhead{$\nu_{min}$ (Hz) } & \colhead{$\bar{\tau}$} & \colhead
{$\Delta E / E_T$ ($\%$)}
}
\startdata
100  &  2 s  & 84 \\
10  &  1000 s  & 96 \\ 
1  &  5.26 d  & 99 \\
\enddata
\end{deluxetable}

\clearpage
\begin{deluxetable}{lcccc}
\tabletypesize{\scriptsize}
\tablecaption{Expected $\Omega_{min}$ of different interferometer pairs for a flat
  background spectrum and an integration time T = 1 year, 
a detection rate $\alpha=90 \%$  and a false alarm rate $\gamma=10\%$.
LHO and LLO stand for LIGO Hanford Observatory and  LIGO Livingston Observatory \label{tbl-sensitivity}}
\tablewidth{0pt}
\tablehead{
  &\colhead{LHO-LHO} & \colhead{LHO-LLO} & \colhead{LLO-VIRGO} & \colhead{VIRGO-GEO}
}
\startdata
initial& $4 \times 10^{-7}$ & $4 \times 10^{-6}$ &  $8 \times 10^{-6}$
&  $8 \times 10^{-6}$  \\
advanced & $6 \times 10^{-9}$  & $1 \times 10^{-9}$  & &  \\
\enddata
\end{deluxetable}

\clearpage

\clearpage

\clearpage
\begin{figure}
\plotone{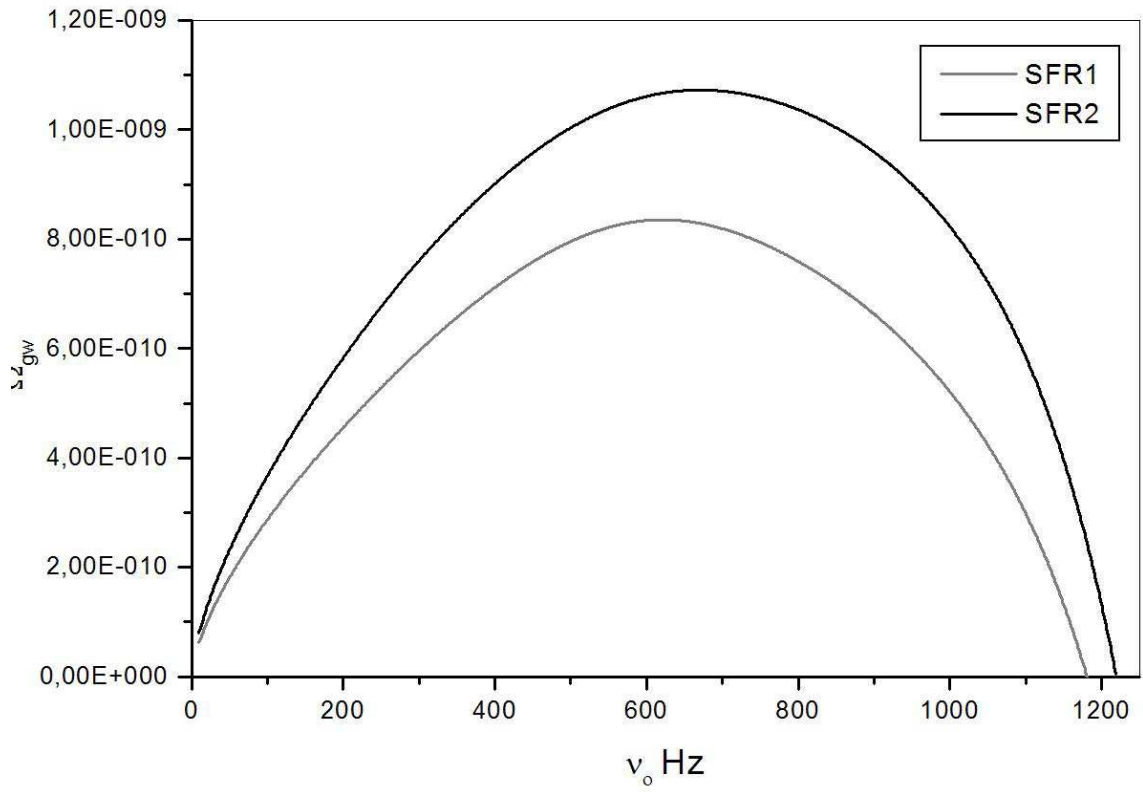}
\caption{Spectrum of the expected gravitational energy density
  parameter $\Omega_{gw}$ for the continuous regime. Results are shown
for the two star formation rates adopted in this work.
\label{fig-spectrum}} 
\end{figure}

\clearpage

\clearpage

\clearpage

\end{document}